# Solution of the Path Integral for the Quantum Pendulum

I. H Duru *
Izmir Institute of Technology
Izmir, Turkey

*e-mail: ihduru@galois.iyte.edu.tr

**Abstract**

Path integration for the potential V=αcosθ is performed. Satisfaction of the corresponding Schrödinger equation by the resulting Feynman kernel is demonstrated. Expressions for the related Green function are presented.

## I. Introduction

Starting from the first examples of linear and quadratic one dimensional potentials by Feynman [1], path integral solutions are available for almost all quantum mechanical problems [2]. One common technique for path integral solutions is the application of a suitable canonical transformation to convert the problem in hand into a soluble, essentially a quadratic type potential. The H-atom [3], and Morse potential [4] are the examples to these techniques. One other common method is the connecting the problem to the motion in a group manifold. For example there is a group of potential related to SU(2) group [5]. Pöschl-Teller and Wood-Saxon are of these type of potentials [5, 6].

The only remaining unsolved problem is the quantum pendulum, that is V=αcosθ potential. It is neither connected to a quadratic potential nor related to the motion over any group manifold. The physical pendulum in quantum mechanics was first discussed in 1928 by Condon [7]. The time independent Schrödinger equation of the problem is of Mathieu differential equation type [8]:

$$\frac{d^2 y}{d\theta^2} + (a + 2q\cos\theta)y = 0 \qquad (1)$$

y is the Mathieu function, that is the elliptic cylindrical functions type which are expressible as an infinite series in terms of sinusoidal or Bessel functions. The expansion coefficients are obtained through very complicated recursion formulas [8].

In this note quantum pendulum is treated by path integration. The Feynman kernel and Green function are derived.

**II. Feynman Kernel**

Path integral for the potential

$$V = \alpha \cos\theta, \qquad \theta: 0 \to 2\pi, \qquad (1)$$

with $\hbar = c = 1$, is

$$\kappa(\theta_a, \theta_b; T = t_b - t_a) = \int D\theta \, e^{i \int_0^T dt (\frac{\mu}{2}\dot\theta^2 - \alpha\cos\theta)}$$

which stands for the usual time-sliced form [1]

$$\kappa = \lim_{\substack{\varepsilon \to 0 \\ n \to \infty}} (\prod_{j=1}^{n} d\theta_j) \prod_{j=1}^{n+1} \left( \sqrt{\frac{i\mu}{2\pi\varepsilon}} \, e^{i\varepsilon \frac{\mu}{2}\left(\frac{d\theta_j}{dt}\right)^2 - i\varepsilon\alpha\cos\theta_j} \right) \qquad (2)$$

with $T = t_b - t_a = (n+1)\varepsilon; \qquad \frac{d\theta_j}{dt} = \frac{1}{\varepsilon}(\theta_j - \theta_{j-1}); \qquad \theta_0 = \theta_a; \theta_{n+1} = \theta_b.$

In the Hamiltonian formulation the kernel (2) is

$$\kappa = \sum_{N=-\infty}^{\infty} \left( \int_{-\infty}^{\infty} \prod_{j=1}^{n} d\theta_j \right) \prod_{j=1}^{n+1} \left( \frac{dp_j}{2\pi} e^{ip_j(\theta_j - \theta_{j-1}) - i\varepsilon \frac{p_j^2}{2\mu} - i\varepsilon\alpha\cos\theta_j} \right) e^{2\pi i N p_{n+1}} \qquad (3)$$

Note that by shifting $\theta_b \to \theta_b + 2\pi N$ together with the change of the range of the interval over $d\theta_j$ as $\theta_j = -\infty \to \infty$, one takes care of the periodicity of the potential (1) [9]. With the representation formula of $\delta$-function,

$$\sum_{N=-\infty}^{\infty} e^{2i\pi N p_{n+1}} = \sum_{N=-\infty}^{\infty} \delta(p_{n+1} - N) \qquad (4)$$

(3) can be rewritten as

$$\kappa = \sum_{N=-\infty}^{\infty} \int_{-\infty}^{\infty} \frac{dp_{n+1}}{2\pi} \delta(p_{n+1} - N)$$

$$x \prod_{j=1}^{n} (\int_{-\infty}^{\infty} \frac{d\theta_j}{2\pi} dp_j e^{i\theta_j(p_j - p_{j+1})})(e^{ip_{n+1}\theta_b - ip_1\theta_a}) \prod_{j=1}^{n+1} \left( e^{-i\varepsilon \frac{p_j^2}{2\mu} - i\varepsilon\alpha\cos\theta_j} \right) \qquad (5)$$

Using the well-known Jacobi-Anger expansion [7] the potential in (5) can be written in terms of the Bessel functions as

$$e^{i\varepsilon\alpha\cos\theta_j} = \sum_{m_i=-\infty}^{\infty} i^{m_i} e^{im_i\theta_j} J_{m_i}(-\alpha\varepsilon) \qquad (6)$$

With this expansion (5) becomes

$$\kappa = \sum_{N=-\infty}^{\infty} \int_{-\infty}^{\infty} \frac{dp_{n+1}}{2\pi} \delta(p_{n+1} - N)$$

$$x \sum_{m_i=-\infty}^{\infty} \left\{ e^{i(p_{n+1}+m_{n+1})\theta_b - ip_1\theta_a} \prod_{j=1}^{n} (\int_{-\infty}^{\infty} dp_j \int_{-\infty}^{\infty} \frac{d\theta_i}{2\pi} e^{i\theta_i(p_j - p_{j+1} + m_j)}) \prod_{j=1}^{n+1} (e^{-i\varepsilon \frac{p_j^2}{2\pi}} i^{m_i} J_{m_i}(-\varepsilon\alpha)) \right\}$$

(7)

Following the integrations over $\prod_{j=1}^{n} d\theta_j$ which lead the $\delta$-functions $\prod_{j=1}^{n} \delta(p_j - p_{j+1} + m_j)$ one simply integrates over $\prod_{j=1}^{n} dp_j$ to get

$$p_i = p_{i+1} - m_i \qquad (8)$$

and obtains (with $N + m_{n+1} = M$ and $\varepsilon = T/(n+1)$)

$$\kappa = \sum_{M=-\infty}^{\infty} \lim_{n\to\infty} \frac{1}{2\pi} \sum_{m_i=-\infty}^{\infty} e^{iM(\theta_b-\theta_a)+i(\sum_l^{n+1} m_l)\theta_a}$$

$$x \left\{ e^{-i\frac{T}{2\mu(n+1)}\sum_j^{n+1}(M-\sum_{l=j}^{n+1} m_l)^2} \prod_{j=1}^{n+1} i^{m_j} \mathcal{J}_{m_j}\left(-\alpha \frac{T}{n+1}\right) \right\} \quad (9)$$

Summations over $m_j$ can be performed by making use of the expansion formula [8]

$$\mathcal{J}_l(z+s) = \sum_{n=-\infty}^{\infty} \mathcal{J}_{l-r}(s)\mathcal{J}_r(z), \qquad |s| > |z| \quad (10)$$

Consider $m_n$ and $m_{n+1}$. Defining $k_n$ by

$$m_n + m_{n+1} = k_n, \qquad m_n = k_n - m_{n+1}$$

the related factor in (9) becomes

$$\sum_{m_{n+1}=-\infty}^{\infty} e^{-i\frac{T}{2\mu(n+1)}(\mu^2+\sum_{j=1}^{n}[M-(k_n+\sum_{l=i}^{n-1} m_l)]^2)-i\frac{T}{2\mu(n+1)}(-2Mm_{n+1}+m_{n+1}^2)+i(k_n+\sum_{l=1}^{n-1} m_l)\theta_a}$$

$$\times e^{k_n} \mathcal{J}_{k_n-m_{n+1}}\left(-\alpha \frac{T}{n+1}\right) \mathcal{J}_{m_{n+1}}\left(-\alpha \frac{T}{n+1}\right)$$

$$= \sum_{k_n=-\infty}^{\infty} e^{-i\frac{T}{2\mu(n+1)}\left(\mu^2+\sum_{i=1}^{n}\left[M-(k_n+\sum_{\substack{l=j\\ \neq n}}^{n-1} m_l)\right]^2\right)+i(k_n+\sum_{l=1}^{n-1} m_l)\theta_a} i^{k_n}\mathcal{J}_{k_n}\left(-\alpha \frac{2T}{n+1}\right) \quad (11)$$

with the vanishing $n \to \infty$ limit of the term in the exponent due to the unpaired $m_{n+1}$:

$$\lim_{n\to\infty} -i\frac{T}{2\mu(n+1)}(-2\mu m_{n+1} + m_{n+1}^2) = 0 \qquad (12).$$

In the following applications of the formula (10) one does not need any limiting processes as above, for all the remaining $m_j$'s appear with paired indices.
In the second step summation over $m_{n-1}$ is performed by defining

$$m_{n-1}+k_n=k_{n-1}, \qquad k_n=k_{n-1}-m_{n-1}$$

The related factor become

$$\sum_{k_{n-1}=-\infty}^{\infty} e^{-i\frac{T}{2\mu(n+1)}\left(M^2+\sum_{j=1}^{n}\left[M-(k_{n-1}+\sum_{\substack{l=j\\\neq n,n-1}}^{n-2} m_l)\right]^2\right)+i(k_{n-1}+\sum_{l=1}^{n-2} m_l)\theta_a} i^{k_{n-1}} J_{k_{n-1}}\left(-\alpha\frac{3T}{n+1}\right)$$
(13)

In the next step $m_{n-2}$ summation is done by defining

$$m_{n-2}+k_{n-1}=k_{n-2}, \qquad k_{n-1}=k_{n-2}-m_{n-2}.$$

In the final step $m_1$ summation is performed to obtain

$$\sum_{k_1=-\infty}^{\infty} e^{-i\frac{T}{2\mu(n+1)}(M^2+\sum_{j=1}^{n}[M-k_1]^2)+ik_1\theta_a} i^{k_1} J_{k_1}\left(-\alpha\frac{(n+1)T}{n+1}\right) \qquad (14)$$

The kernel of (9) then takes its final form (with $k \equiv k_1, L \equiv M - k_1$)

$$K = \sum_{L=-\infty}^{\infty} \lim_{n\to\infty} \frac{1}{2\pi} \sum_{k=-\infty}^{\infty} e^{-i\frac{T}{2\mu(n+1)}((L+k)^2+nL^2)} i^k J_k(-\alpha T) e^{iL(\theta_b-\theta_a)+ik\theta_b} \qquad (15)$$

or with

$$\lim_{n\to\infty} e^{-i\frac{T}{2\mu(n+1)}(2Lk+k^2)} = 1$$

one arrives at

$$\kappa(\theta_a, \theta_b; T) = \sum_{L,k=-\infty}^{\infty} \frac{1}{2\pi} e^{-i\frac{L^2}{2\mu}T} i^k \mathcal{J}_k(-\alpha T) e^{iL(\theta_b-\theta_a)+ik\theta_b} \quad (16)$$

To make the initial and final points separated one can employ the expansion formula (16) to get

$$\kappa(\theta_a, \theta_b; T) = \sum_{L,k=-\infty}^{\infty} \frac{1}{2\pi} \sum_{r=-\infty}^{\infty} e^{-i\frac{L^2}{2\mu}(t_b-t_a)} i^k \mathcal{J}_{k-r}(-\alpha t_b) \mathcal{J}_r(\alpha t_a) e^{i(L+k)\theta_b - iL\theta_a} \quad (17)$$

For $\alpha=0$ only k=0 survives. Thus (16) becomes

$$\kappa(\alpha = 0) = \frac{1}{2\pi} \sum_{L=-\infty}^{\infty} e^{-i\frac{L^2}{2\mu}T} e^{iL(\theta_b-\theta_a)}$$

which is the free rotation kernel.

### III. Satisfaction of the Schrödinger Equation

To demonstrate that the Schrödinger equation

$$\left(-\frac{1}{2\mu}\frac{\partial^2}{\partial \theta_b^2} - \alpha\cos\theta_b - i\frac{\partial}{\partial t_b}\right)\kappa = 0 \quad (18)$$

is satisfied, write the potential as

$$\alpha\cos\theta_b \kappa = \frac{\alpha}{2}\left(e^{i\theta_b} + e^{-i\theta_b}\right)\kappa \quad (19)$$

The first term is

$$\frac{\alpha}{2}e^{i\theta_b}K = \frac{\alpha}{2}\frac{1}{2\pi}\sum_{L,k=-\infty}^{\infty} e^{-i\frac{L^2}{2\mu}T} i^k J_k(-\alpha T)e^{iL(\theta_b-\theta_a)+i(k+1)\theta_b}$$

which by shifting k as k→k-1 becomes

$$\frac{\alpha}{2}e^{i\theta_b}K = \frac{\alpha}{2}\frac{1}{2\pi}\sum_{L,k=-\infty}^{\infty} e^{-i\frac{L^2}{2\mu}T} i^{k-1} J_{k-1}(-\alpha T)e^{iL(\theta_b-\theta_a)+ik\theta_b} \quad (20)$$

The second term of (19) is handled by the shift k→k+1 to obtain an expression similar to the one in (20) with k-1 replaced by k+1. Using the property of Bessel functions [8]

$$J_{k-1}(z) - J_{k+1}(z) = 2\frac{d}{dz}J_k(z) \quad (21)$$

with

$$\frac{d}{dz}J_k(z=-\alpha T) = -\alpha\frac{d}{dt_b}J_k(-\alpha T), \qquad T=t_b-t_a \quad (22)$$

one has

$$\alpha\cos\theta_b K = \frac{i}{2\pi}\sum_{L,k=-\infty}^{\infty} e^{-i\frac{L^2}{2\mu}T} i^k \frac{d}{dt_b}J_k(-\alpha T)e^{iL(\theta_b-\theta_a)+ik\theta_b} \quad (23)$$

It is then a trivial matter to see the satisfaction of (18), for both L and k summations are from -∞ to +∞.

To show that the Schrödinger equation is satisfied at $\theta_a$, ta, one has to make shifts L→L±1 simultaneously with k→k∓1.

## IV. Green Function

The Green function

$$G(Q_a, Q_b; E) = \int_0^\infty dT\, e^{iET} \kappa(Q_a, Q_b; T) \tag{24}$$

can be expressed in alternate representations.

(i) Employ the series representation of Bessel function in $\kappa$ [8]

$$\mathcal{J}_k(-\alpha T) = \sum_{l=0}^{\infty} (-)^l \frac{1}{l!\Gamma(k+l+1)} \left(-\alpha \frac{T}{2}\right)^{k+2l} \tag{25}$$

Making use of the identity

$$\int_0^\infty dx\, e^{i\Lambda x} x^s = \lim_{\delta \to 0} (-i)^s \frac{\partial}{\partial \delta^s} \int_0^\infty dx\, e^{i(\Lambda+\delta)s} \tag{26}$$

One can perform the integration over dT in (24) to obtain

$$G(\theta_a, \theta_b; E) = \frac{1}{2\pi} \sum_{L,k=-\infty}^{\infty} \sum_{l=0}^{\infty} \frac{(-)^l}{l!\Gamma(k+l+1)} \left(\frac{\alpha}{2}\right)^{k+2l} \frac{e^{iL(\theta_b-\theta_a)+ik\theta_b}}{\left(E-\frac{L^2}{2\mu}\right)^{1+k+2l}} \tag{27}$$

or

$$G(\theta_a, \theta_b; E) = \frac{1}{2\pi} \sum_{L,k=-\infty}^{\infty} \frac{1}{\left(E-\frac{L^2}{2\mu}\right)} \mathcal{J}_k\left(\frac{\alpha}{2(E-\frac{L^2}{2\mu})}\right) e^{iL(\theta_b-\theta_a)+ik\theta_b} \tag{28}$$

(ii) To have an alternative expression for G, the integral representation can be used for the Bessel function [8]

$$\mathcal{J}_k(-\alpha T) = \frac{1}{2\pi} \int_{-\pi}^{\pi} d\vartheta\, e^{-ik\vartheta - i\alpha T \sin\vartheta} \tag{29}$$

One then obtains

$$G(\theta_a, \theta_b; E) = \frac{1}{2\pi}\sum_{L,k=-\infty}^{\infty} \frac{1}{2\pi}\int_{-\pi}^{\pi} d\vartheta \frac{e^{-ik\vartheta}}{E-\frac{L^2}{2\mu}-\alpha\sin\vartheta} e^{iL(\theta_b-\theta_a)+ik\theta_b} \quad (30)$$

(iii) Employing the integration formula of Bessel function and the exponential [8], given as hypergeometric function

$$\int_0^\infty dx\, x^{\mu-1} e^{-\Lambda x} J_\nu(\beta x) = \frac{\left(\frac{\beta}{2\Lambda}\right)^\nu \Gamma(\nu+\mu)}{\Lambda^\mu \Gamma(\nu+1)} F\left(\frac{\nu+\mu}{2}, \frac{\nu+\mu+1}{2}; \nu+1; -\frac{\beta^2}{\Lambda^2}\right) \quad (31)$$

One can express the Green function as

$$G(\theta_a, \theta_b; E) = \frac{1}{2\pi}\sum_{L,k=-\infty}^{\infty} \frac{1}{\frac{L^2}{2\mu}-E}\left[\frac{-i\alpha}{2\left(\frac{L^2}{2\mu}-E\right)}\right]^k F\left(\frac{k+1}{2}, \frac{k}{2}+1; k+1; -\frac{\alpha^2}{\left(\frac{L^2}{2\mu}-E\right)^2}\right) e^{iL(\theta_b-\theta_a)+ik\theta_b}$$

(32)

## V. Summary and Discussion

Feynman kernel for the quantum pendulum is obtained as an infinite series of plane waves with time dependent Bessel functions being the expansion coefficients. Expressions for the corresponding Green function are also obtained.

Although the Feynman kernel of (16) is very compact, its time dependence is through the Bessel functions which is unusual. This peculiar time dependence is certainly related to the very complicated continued fraction expression of the energy eigenvalues of the time-independent Schrödinger equation of the problem, that is the Mathieu differential equation [10]. Note that a similar peculiarity exists for the linear potential problem; for which the Feynman kernel is very compact [1], while the wave functions are the Airy functions [2] and the energy eigenvalues are given as continued fractions.

Finally note that the above mentioned unusual time dependence of the quantum pendulum kernel should be connected to a property of the corresponding classical problem: The classical pendulum is solved by Jacobi elliptic function with one usual real period and one imaginary period. The imaginary period corresponds to the "swing through the top point $\theta = \pi$" [11].

**Acknowledgement**

The author thanks D. A. Demir, C. Saclioglu and T. Senger for discussions and F. Duru for critical reading of the manuscript.